%% file: deltaphi.tex
\documentclass[11pt,twoside]{myArticle}       
\usepackage{amsmath,amssymb,amsfonts} 
\usepackage{graphicx}
\usepackage{epsfig}
\usepackage{color}                    
\usepackage{hyperref}                 
\usepackage{myFancyhdr}               
\usepackage{makeidx}                  
\usepackage{helvet}                   
\usepackage{setspace}                 
\usepackage{fancybox}                 
\usepackage{float}                    
\usepackage{xspace}                   
\usepackage{wasysym}                  
\usepackage[numbers,sort&compress]{natbib}
\usepackage{subfigure}
\usepackage{atlasphysics}
\usepackage{ourphysics}
\usepackage{amsmath}
\usepackage{amssymb}
\usepackage{atlas_title,ifthen}
\usepackage{mathptmx}
\usepackage{graphics}
\usepackage{helvet}
\usepackage{subfigure}
\usepackage{rotating}                 
\usepackage{atlasphysics}             
\newlength{\figwidth}
\newcommand{\icaption}[2][!*!,!]{\hspace*{\capindent}%
  \begin{minipage}{\capwidth}
    \ifthenelse{\equal{#1}{!*!,!}}%
      {\caption{#2}}%
      {\caption[#1]{#2}}
      \vspace*{3mm}
  \end{minipage}}
\definecolor{darkblue1}{rgb}{0,0,.7}
\definecolor{darkblue}{rgb}{0,0,.3}
\definecolor{darkred}{rgb}{0.5,0,0}
\pagecolor{white} 
\color{black}     
\bibstyle{plain}
\input MyDef

\begin{document}

\newpage
       
\begin{titlepage}

\vspace{-1cm}
\begin{flushright}
{\sf\em IRFU-10-17} \\
{\sf\em \today} \\
\end{flushright}
\def\UrlFont{\rm}

\def\miniPageOffset{0.2cm}
\def\miniPageWidth{13.5cm}
\vspace*{\stretch{20}}
\begin{flushleft}
\hspace{\miniPageOffset}\begin{minipage}{\miniPageWidth}
{\sf\Huge\bfseries\boldmath Z boson transverse momentum spectrum from the lepton angular distributions}
\end{minipage}
\end{flushleft}
\vspace{2.0cm}
\begin{flushright}
{\sf\Large  M.~Boonekamp (CEA), } 
{\sf\Large  M.~Schott (CERN)} 
\vspace*{\stretch{1}}
\end{flushright}

\def\UrlFont{\sf}
\noindent\rule[1ex]{\textwidth}{1pt}
{\small\sf
{\sf\bfseries Abstract} --- 
In view of recent discussions concerning the possibly limiting
energy resolution systematics on the measurement of the Z boson
transverse momentum distribution at hadron colliders, we propose a novel measurement method based on
the angular distributions of the decay leptons. We also introduce a
phenomenological parametrization of the transverse momentum distribution that adapts
well to all currently available predictions, a useful tool to quantify
their differences.

}
\noindent\rule[1ex]{\textwidth}{1pt}

\begin{center}
{\small\sf

}
\end{center}

\thispagestyle{empty}
\newpage

\end{titlepage}


\newpage
%

\section{Introduction\label{sec:intro}}
\input{intro.tex}

%

\section{A parametrized form for the \pt\ distribution \label{param}}
\input{parametrization.tex}

%

\section{\pt\ spectrum from the angular distributions \label{deltaphi}}
\input{ptspectrum.tex}

%

\subsection{\label{ExpectedPrecision} Expected Precision}
\input{expectedPrecision.tex}

\subsection{\label{Generators} Comparison of different Monte Carlo Generators}
\input{comparisonMC.tex}

\section{Perspectives \label{perspectives}}
\input{perspectives.tex}

%

\section{Acknowledgements \label{acknowledgement}}
\input{acknowledgements.tex}

%

\end{document}

%% file: MyDef.tex
\renewcommand{\headrulewidth}{0.5pt}    
\renewcommand{\footrulewidth}{0.5pt}    

\newcommand\allFontSize{\small}

\usepackage[bf]{caption}

\newcommand\detailsSize{\allFontSize}
{\begin{myquote}\detailsSize}{\end{myquote}}

\newfloat{codeexample}{H}{loc}
\floatname{codeexample}{\sf\allFontSize Code Example}

\newlength{\susyboxwidth}
\definecolor{DarkGray}{rgb}{0.4,0.42,0.45}
\definecolor{LightGray}{rgb}{0.95,0.96,0.96}

{\VerbatimEnvironment\normalsize
\setlength{\susyboxwidth}{\textwidth}\addtolength{\susyboxwidth}{-2.3\fboxsep}%
\begin{Sbox}\begin{minipage}[t]{\susyboxwidth}\begin{Verbatim}}%
{\end{Verbatim}\end{minipage}\end{Sbox}\vspace{1ex} \fcolorbox{DarkGray}{LightGray}{\TheSbox}}

\newfloat{option}{H}{loc}
\floatname{option}{\sf\allFontSize Option Table}

\usepackage{tabularx}
{\setlength{\susyboxwidth}{\textwidth}\addtolength{\susyboxwidth}{-2.3\fboxsep}%
\begin{Sbox}\begin{tabular*}{\susyboxwidth}{>{\rule[-0.5ex]{.0ex}{3.0ex}\tt\small}p{3cm}>{\tt\small}p{4.5cm}>{\small}p{5.7cm}}
&&\\[-0.7cm]
\rm\small Option  & \rm\small Values    & \rm\small Description\\[0.1cm]\hline
&&\\[-0.45cm]}%
{\\[-0.1cm]\end{tabular*}\end{Sbox}\vspace{1ex} \fbox{\TheSbox}}

{\setlength{\susyboxwidth}{\textwidth}\addtolength{\susyboxwidth}{-2\fboxsep}%
\begin{Sbox}\begin{tabular*}{\susyboxwidth}{>{\rule[-0.5ex]{.0ex}{3.0ex}\tt\small}p{4.5cm}>{\tt\small}p{3.2cm}>{\small}p{5.5cm}}
&&\\[-0.7cm]
\rm\small Option  & \rm\small Values    & \rm\small Description\\[0.1cm]\hline
&&\\[-0.45cm]}%
{\\[-0.1cm]\end{tabular*}\end{Sbox}\vspace{1ex} \fbox{\TheSbox}}

{\setlength{\susyboxwidth}{\textwidth}\addtolength{\susyboxwidth}{-2\fboxsep}%
\begin{Sbox}\begin{tabular*}{\susyboxwidth}{>{\rule[-0.5ex]{.0ex}{3.0ex}\tt\small}p{4.0cm}>{\tt\small}p{3.0cm}>{\small}p{6.2cm}}
&&\\[-0.7cm]
\rm\small Option  & \rm\small Values    & \rm\small Description\\[0.1cm]\hline
&&\\[-0.45cm]}%
{\\[-0.1cm]\end{tabular*}\end{Sbox}\vspace{1ex} \fbox{\TheSbox}}

\newfloat{programs}{H}{loc}
\floatname{programs}{\sf Table}

\usepackage{tabularx}
{\setlength{\susyboxwidth}{\textwidth}\addtolength{\susyboxwidth}{-2\fboxsep}%
\begin{Sbox}\begin{tabular*}{\susyboxwidth}{>{\rule[-0.5ex]{.0ex}{3.0ex}\tt\small}p{4.1cm}>{\small}p{9.5cm}}
&\\[-0.7cm]
\rm\small Macro & \rm\small Description\\[0.1cm]\hline
&\\[-0.45cm]}%
{\\[-0.1cm]\end{tabular*}\end{Sbox}\vspace{1ex} \fbox{\TheSbox}}



%% file: intro.tex
\noindent The transverse momentum distribution of heavy particles at
hadron colliders is a longstanding subject, first discussed in the
context of QCD. Formalisms were developed to predict this
distribution, based on analytical or numerical methods (soft gluon
resummation
theory~\cite{Collins:1984kg,hep-ph/9311341,hep-ph/9704258,hep-ph/0212159},
or parton shower Monte Carlo
programs~\cite{hep-ph/0603175,hep-ph/0210213} respectively). In both
cases, the shape of the 
distribution is predicted qualitatively, but the full result depends
on a limited number of free parameters which need to be extracted from 
measurement. \\

\noindent The distribution is of physical interest for many
reasons. Firstly, the measurement of the overall vector boson
production cross sections is considered an important test of
perturbative QCD, theoretical predictions now being available up to 
NNLO~\cite{hep-ph/0609070}. In the context of the total cross section
measurement, the kinematic cuts imposed on the decay leptons,
reflecting the detector geometric acceptance, require that the observed event rate be corrected by a factor compensating this loss of
acceptance. The fraction of lost events must be precisely controled so 
that the final result contains no significant bias. This in turn
implies that the lepton kinematic distributions, and hence the vector
boson ones from which they derive, need to be known both inside and
outside the selected region.\\

\noindent Another application is the precise measurement of the W
boson mass~\cite{arXiv:0805.2093}. The decay lepton transverse momentum distribution,
or the W boson transverse mass distribution from which this
fundamental parameter is extracted, is a complicated quantity resulting in
part from the W boson transverse momentum spectrum, \dswdpt. The
increasing precision of the measurements of $M_W$ puts ever stronger
constraints on the knowledge of \dswdpt. An important tool for
constraining this distribution is the study of the Z boson transverse
momentum spectrum, \dszdpt.\\

\noindent The recent measurements performed at the Tevatron are
increasingly sensitive to the detector energy resolution, which needs
to be precisely ``subtracted'', or unfolded from the observed
distribution to derive an estimate of the true one. This issue will
become much more important at the LHC, given the high expected
statistics. An alternative variable, $a_T$, was introduced
recently~\cite{arXiv:0807.4956} as a replacement for \pt. Its advantage is
that it is negligibly sensitive to the energy resolution, while still
a good probe of resummation or parton shower mechanisms. \\

\noindent In this paper, we propose a novel method that shares the
insensitivity of $a_T$ to the energy resolution, while remaining a
true \pt\ measurement. The method is based on the measured angular
distributions of the $Z$ boson decay leptons, which, together with the
well known Z boson mass, are sufficient to extract the
\pt\ distribution. \\

\noindent In the following, we first introduce a convenient
parametrization of the \pt\ distribution. It depends on three
intuitive parameters, and adapts well to the available
predictions. It represents a practical tool to quantify differences
between predictions, as well as for the measurement itself. We then
outline the measurement method, and give examples of its performance
in a simplified form. We conclude with some caveats and perspectives
concerning the use of this method in future measurements.

%% file: parametrization.tex
The parametrization we propose relies on a number of simple
arguments. Consider Z boson production at high energy, and at given
mass and rapidity, so that the parton momentum fractions at the hard
vertex are small and fixed. In the low transverse momentum region, the
repeated gluon emission in the initial state generates a gaussian
transverse momentum distribution. Along both the $x$ and $y$ axes,
this "random walk" leads to a distribution proportional to
$$f(p_{x,y};\sqcd)  \,\, dp_{x,y} \sim e^{-\frac{p^2_{x,y}}{2\sqcd^2}} \,\, dp_{x,y}.$$ 
\noindent The \sqcd\ parameter represents the spread of the \pxy\
distribution after all emissions and, in a naive picture, could be
seen as representing the average number of emitted gluons times their
average transverse momentum: $\sqcd \sim \sqrt{N_g} \times p^g_{x,y}$.
Moving to polar coordinates, the distribution becomes:    
\begin{eqnarray}
f(p_x;\sqcd)  \,\, f(p_y;\sqcd)  \,\, d p_{x}  \,\, d p_{y} &\sim&
  e^{-\frac{p^2_{x}}{2\sqcd^2}} \,\, e^{-\frac{p^2_{y}}{2\sqcd^2}} \,\, d p_x \,\, d p_y \nonumber \\
  &=&  e^{-\frac{\pt^2}{2\sqcd^2}} \,\, p_T \,\, d p_T \,\,
  d \phi \,\, \equiv \,\, g_1(p_T;\sqcd) \,\, d p_T. \nonumber
\end{eqnarray}

after a trivial azimuthal integral. At higher \pt, the shape
is dominated by a power law behaviour representing the parton density
functions (PDFs) and the perturbative matrix element:
$$g_2(p_T;a) \sim 1/\pt^\alpha.$$

The transition between the two descriptions is controled by
a parameter $n$, defined such that $p_T^{match} =
n \times \sqcd$. As for the definition of the Crystal Ball
function~\cite{Gaiser:1982yw}, it satisfies smoothness conditions (the
function and its derivative are continuous). The complete
parametrization is, forgetting an overall normalization factor:
\begin{eqnarray}
\label{eqnpara}
g(p_T; \sqcd, \alpha, n) &=& p_T \,\,
                    e^{-\frac{p_T^2}{2\sqcd^2}}, \,\,\,\,\,\, p_T <
                    n \times \sqcd;\nonumber\\ 
                    &=& p_T \,\, \frac{(\frac{\alpha}{n})^\alpha \,
                    e^{-n^2/2}}{(\frac{\alpha}{n} - n
                    + \frac{p_T}{\sqcd})^\alpha}, \,\,\, p_T >
                    n \times \sqcd; 
\end{eqnarray}

\noindent where the parameters $\alpha$, $n$, and \sqcd\ are all positive 
definite. Fitted to various generator level \sdsdpt\ distributions, 
this parametrisation shows a nice behaviour in the range in
$0<\pt<50$~\GeV. Over a wider range, the agreement slightly
deteriorates, due to the fact that the high \pt\ power law with a
constant power is a crude approximation. Both PDFs and matrix
element's power depend on the scale $Q$ of the process, which is
related to $p_T$. The fit quality could be improved by introducing a
running power law, $\alpha(Q^2)$, at the cost of additional free
parameters.\\

\begin{figure}[htbp]
  \begin{center}
    \subfigure[]{\includegraphics[width=0.56\figwidth]{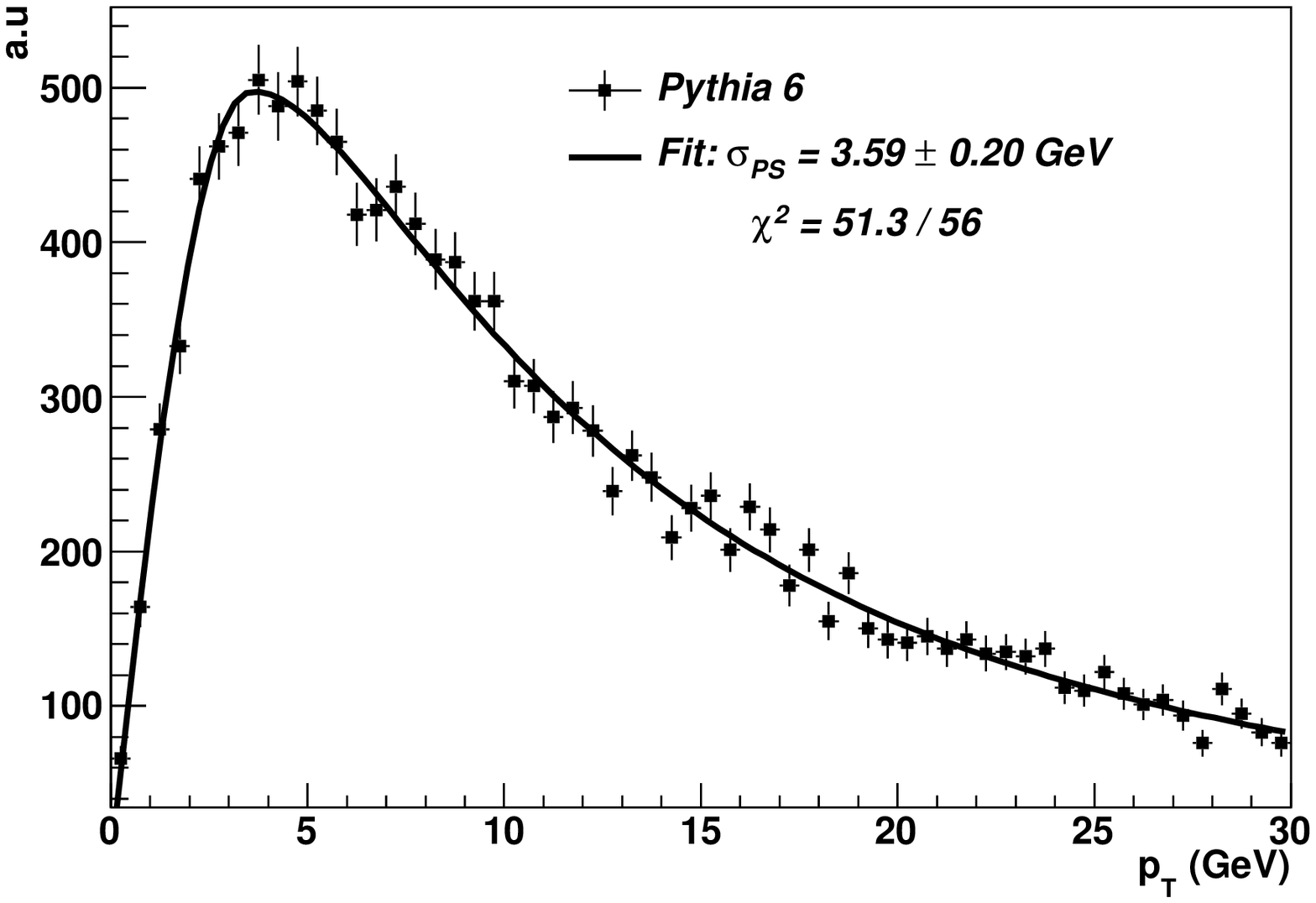}\label{pythia}}
    \subfigure[]{\includegraphics[width=0.56\figwidth]{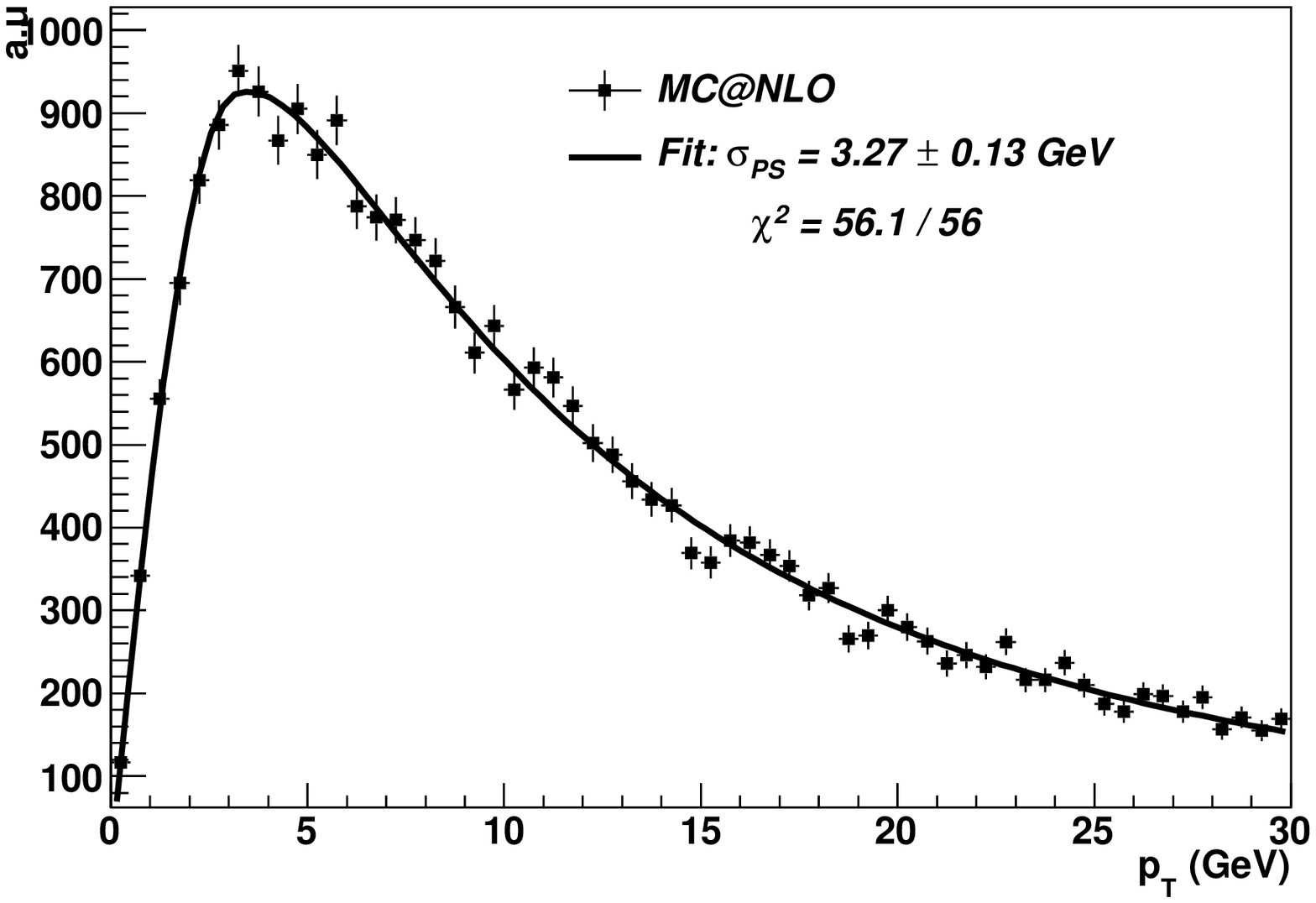}\label{mcatnlo}}
    \subfigure[]{\includegraphics[width=0.56\figwidth]{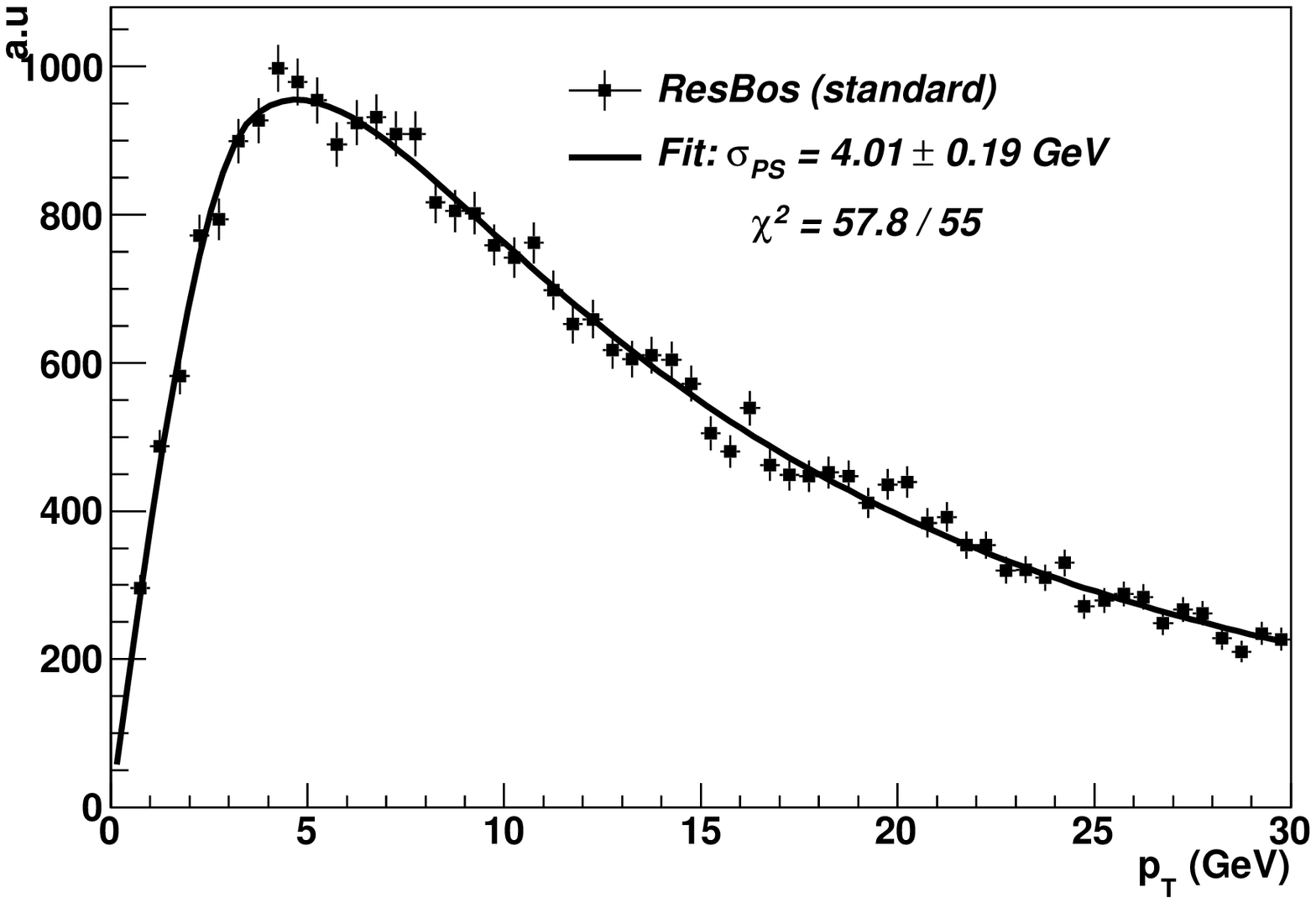}\label{resbosStd1}}
    \subfigure[]{\includegraphics[width=0.56\figwidth]{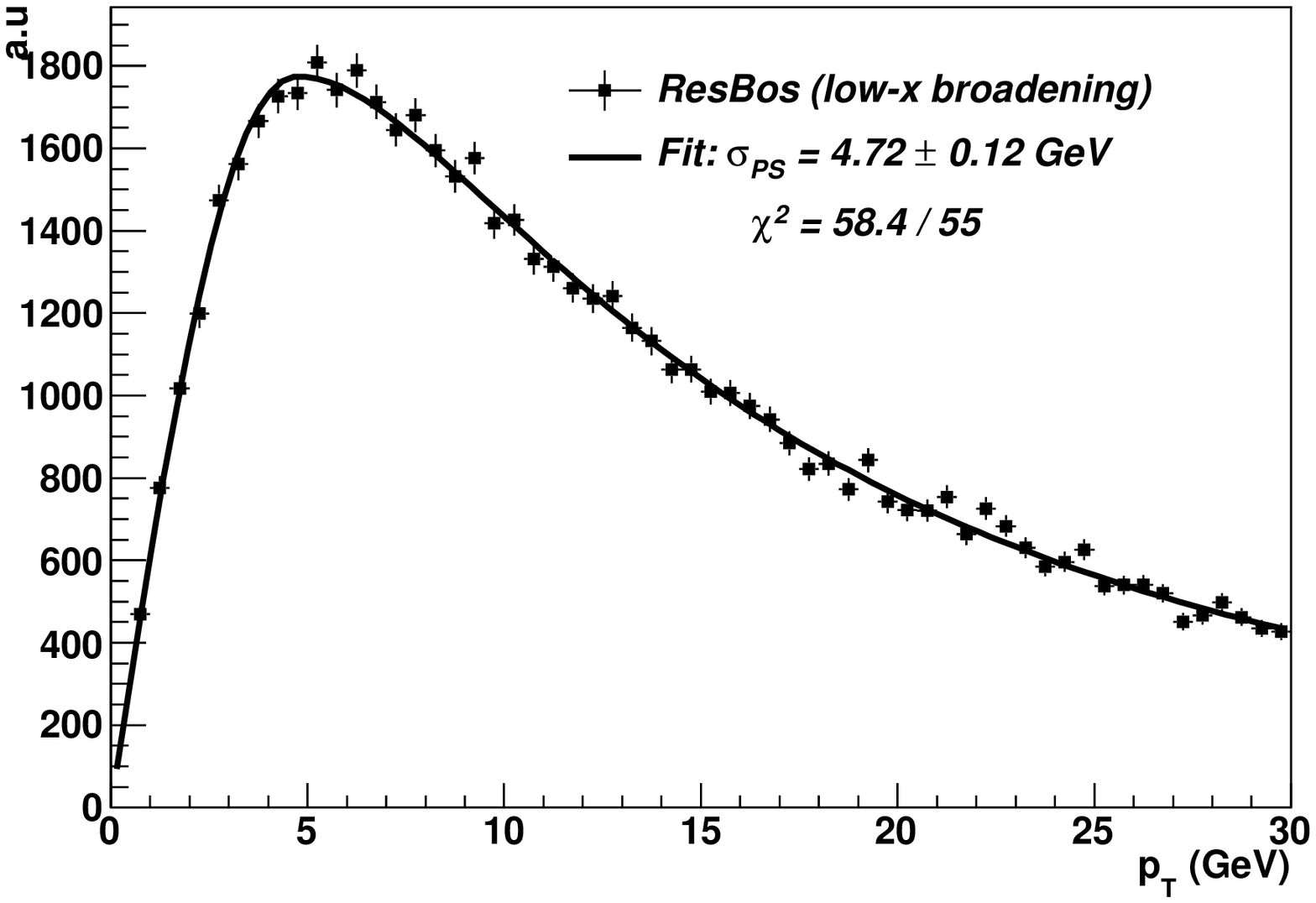}\label{resbosBrd1}}
  \end{center}
  \caption{Generator level \pt\ distribution, at $\sqrt{s} = 14$~TeV,
    $Q=M_Z$ and $y_Z=0$ , as predicted by {\tt
      PYTHIA}~\subref{pythia}, {\tt MC@NLO}~\subref{mcatnlo}, the 
    standard {\tt RESBOS}~\subref{resbosStd1}, and by the version
    including low-$x$ broadening effects~\subref{resbosBrd1}.
  \label{gendep}}
\end{figure}

\noindent Figure~\ref{gendep} shows the parametrisation fitted to
distributions obtained using the Monte Carlo event generators {\tt PYTHIA}~\cite{hep-ph/0603175}, {\tt
  MC@NLO}~\cite{arXiv:0812.0770}, and two versions of {\tt
  RESBOS}~\cite{hep-ph/9311341,hep-ph/9704258,hep-ph/0410375}: the
  default computation, and a computation including small-$x$ 
broadening effects. All distributions are obtained at $\sqrt{s} =
14$~TeV, $Q=M_Z$ and $y_Z=0$. The fit quality is good, with $\chi^2
\sim 1$ in all cases.\\

\begin{figure}[htbp]
  \begin{center}
    \subfigure[]{\includegraphics[width=0.56\figwidth]{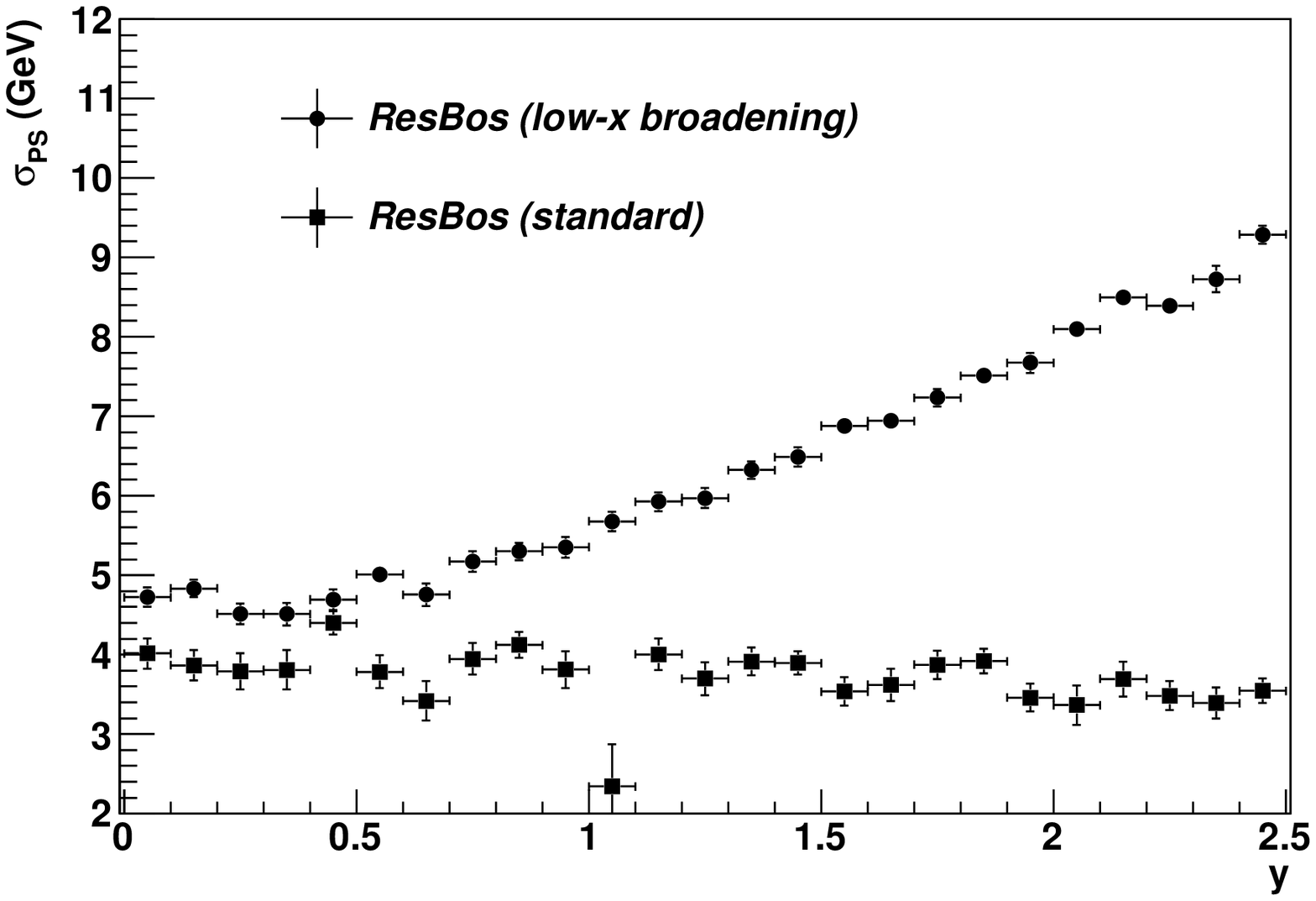}\label{resbosStd2}}
    \subfigure[]{\includegraphics[width=0.56\figwidth]{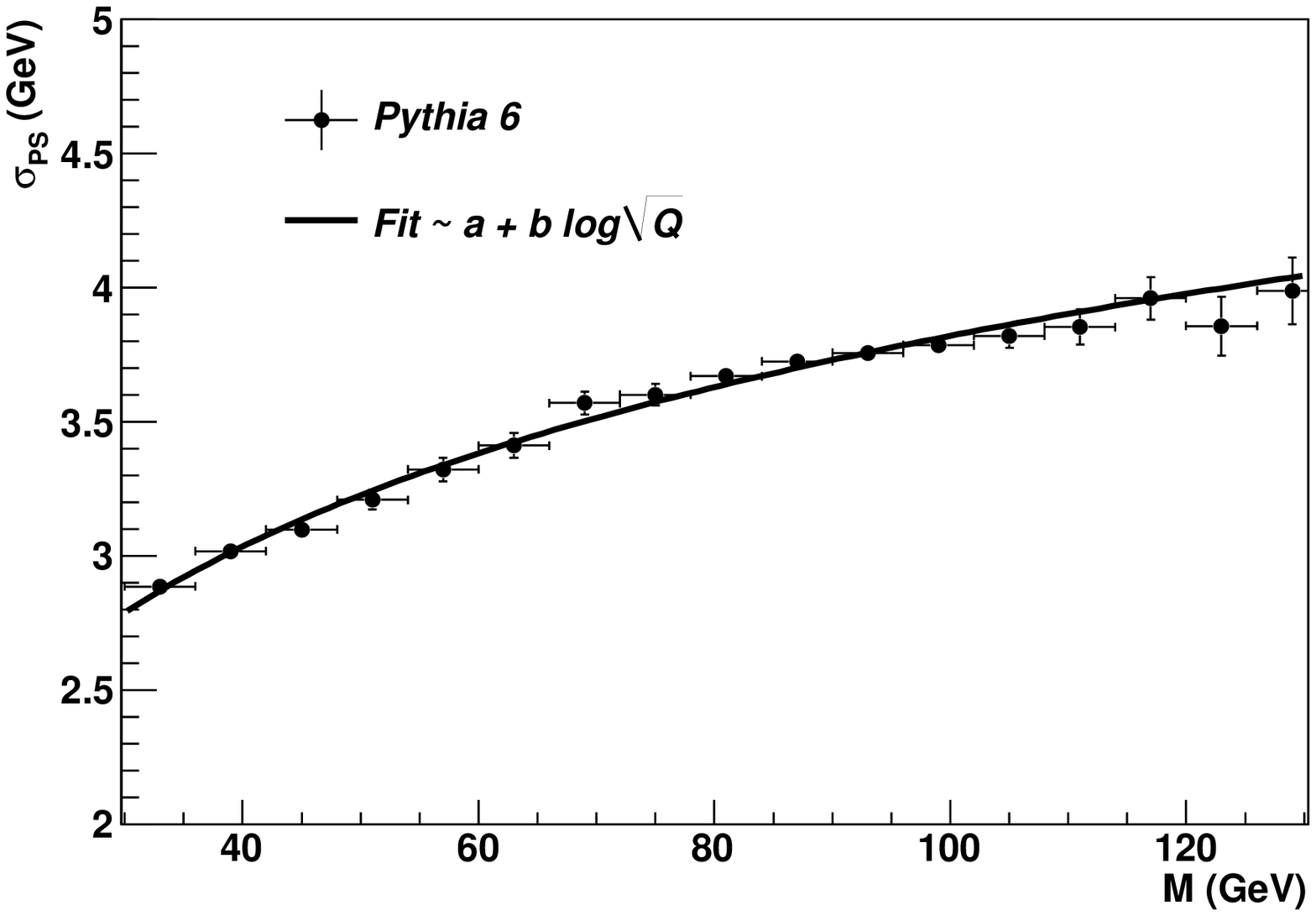}\label{pythia2}}
  \end{center}
  \caption{Rapidity dependence of \sqcd, as predicted by the standard
    {\tt RESBOS}, and by the version including
    low-$x$ broadening effects~\subref{resbosStd2}; mass dependence of \sqcd, as predicted by {\tt PYTHIA}~\subref{pythia2}.
  \label{rapdep}}
\end{figure}

\noindent It is interesting to study the dependence of the
\sqcd\ parameter as a function of rapidity, as illustrated in
Figure~\ref{rapdep}. The standard {\tt RESBOS} prediction shows falling
values of this parameter at higher rapidity, an effect generally
expected from the decreasing phase space on one side of the parton
shower. The modified version, however, shows an increase of this 
parameter, resulting from the low-$x$ effects. It would be interesting
to measure this dependence in the current and forthcoming hadron
collider data.

\noindent Finally, one can study the $Q$ dependence of the
\sqcd\ parameter. As stated above, it is naively proportional to
$\sqrt{N_g}$, the number of gluons emitted in the initial state. $N_g$
is, on the other hand and according to the Altarelli-Parisi evolution
equations~\cite{Gribov:1972ri,Altarelli:1977zs,Dokshitzer:1977sg},
proportional to the logarithm of the scale variation 
between the original proton and the hard process, $N_g \sim
\log(Q^2)$. Plotting \sqcd\ as a function of $Q$, taken to be the
boson invariant mass event by event, shows a behaviour
following $\sqcd \sim \sqrt{\log(Q^2)}$ as expected in this simple picture.

%% file: ptspectrum.tex
\subsection{Methodology}
The proposed measurement procedure is suggested by observing that at
given transverse momentum and at fixed mass, the $Z$ boson angular
distribution can be written as the product of the lepton angular
distribution in the $Z$ rest frame, and a factor relating the lepton
angles in the rest frame and laboraty frame. For simplicity, we
consider the azimuthal angular distribution only: 

\begin{equation}
\left({\frac{d\sigma}{\Delta\phi}}\right)_{p_T} =
\left(\frac{d\sigma}{d\phi^*}\right)_{p_T} \times
\left(\frac{d\phi^*}{d\Delta\phi}\right)_{p_T} 
\label{eq1}
\end{equation}

\noindent Above, $\phi^*$ is the azimuthal angle in the rest frame, and
$\Delta\phi = \phi_1 - \phi_2$ the angular separation in the
laboratory frame. The transverse momentum distribution can be inferred
by noting that the overall $\Delta\phi$ distribution is the integral
of the above over the $p_T$ distribution: 

\begin{equation}
\frac{d\sigma}{d\Delta\phi} = \int
\left(\frac{d\sigma}{\Delta\phi}\right)_{p_T} 
\frac{d\sigma}{dp_T} dp_T
\label{eq2}
\end{equation}

\noindent In Eq.~\ref{eq1}, the first factor on the right hand
side has a well defined form, and can be computed perturbatively. In
the Collins-Soper frame\footnote{Defined as the gauge boson
  rest frame which maximizes the projections of the beam momenta on
  the $z$-axis.}~\cite{Collins:1977iv}, one finds: 

\begin{equation}
\left(\frac{d\sigma}{d\Omega}\right)_{p_T} \sim 1 + \cos^2\theta^* +
(\frac{1}{2}-\frac{3}{2}\cos^2\theta^*) \,\, A_0 +
\cos\theta^*\sin\theta^*\cos\phi^* \,\, A_1 + \frac{1}{2} \sin^2\theta^*\cos 2\phi^*
\,\, A_2
\end{equation}

\noindent where the coefficients $A_i$ can be calculated
perturbatively and are functions of the kinematic variables $s$, $y$, $p_T$. The
second factor is purely kinematic. In the simplest  
case where the system is purely transverse (all rapidities are 0 and
momenta are purely transverse), the
relation between $\phi^*$ and $\Delta\phi$ takes the following form:

\begin{equation}
\phi^* = \cos^{-1}\left( \frac{1}{\beta} \sqrt{\frac{2\beta^2-\cos\Delta\phi}{1-\cos\Delta\phi}} \right)
\end{equation}

\noindent where $\beta = \frac{p_T}{E}$, and from which the the
derivative $\frac{d\phi^*}{d\Delta\phi}$ in Equation~\ref{eq1} can  be computed. According
to the above, the $\Delta\phi$ distribution is 
directly sensitive to $p_T$: small values of $\Delta\phi$ indicate
large $p_T$ values. In the general case, the polar decay angles
complicate the picture significantly, as at finite but modest $p_T$,
small $\Delta\phi$ values are also obtained from forward
decays ($|\eta| \gg 0$), leading to a small projection of the lepton pair 
opening angle in the transverse plane. In addition, the expressions
have to be  integrated over the $Z$ boson lineshape, so that in
practice we have to extract the factors from large Monte Carlo
generated event samples. For the sake of simplicity, we stick to
Eq.~\ref{eq2} and compute the the $\Delta\phi$ distribution at given
$p_T$ integrating over all other relevant variables:

\begin{equation}
\left(\frac{d\sigma}{\Delta\phi}\right)_{p_T} = \int
\left(\frac{d\sigma}{\Delta\phi}\right)_{p_T;m,y,\eta} dm \, dy \, d\eta 
\end{equation}

\noindent For the analysis, we use a sample of $10^6$ events,
generated with {\tt MC@NLO}. While this approximation is not optimal
as the lepton rapidities are also measured, providing additional
information which is not exploited here, it is sufficient for the
purpose of demonstration. Statistical sensitivities discussed here
should thus be understood as conservative.\\

\begin{figure}[tb]
\centering
  \begin{minipage}[b]{0.56\figwidth}
  \includegraphics[width=0.56\figwidth,angle=0]{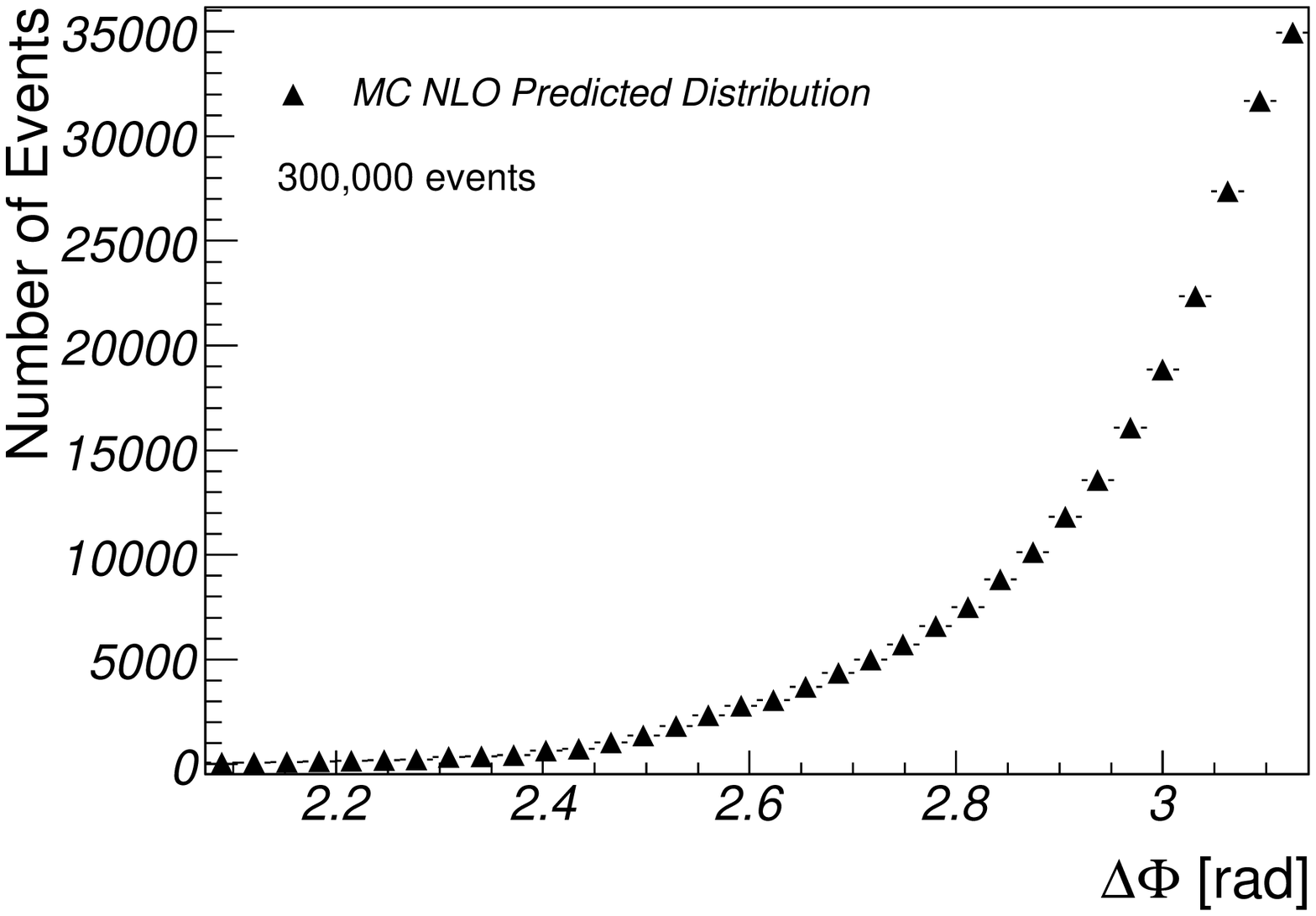}
  \caption{\label{labelAziFull}Opening-angle distribution for two Z boson decay
    leptons.}  
  \end{minipage}
  \begin{minipage}[b]{0.56\figwidth}
    \includegraphics[width=0.56\figwidth, angle=0]{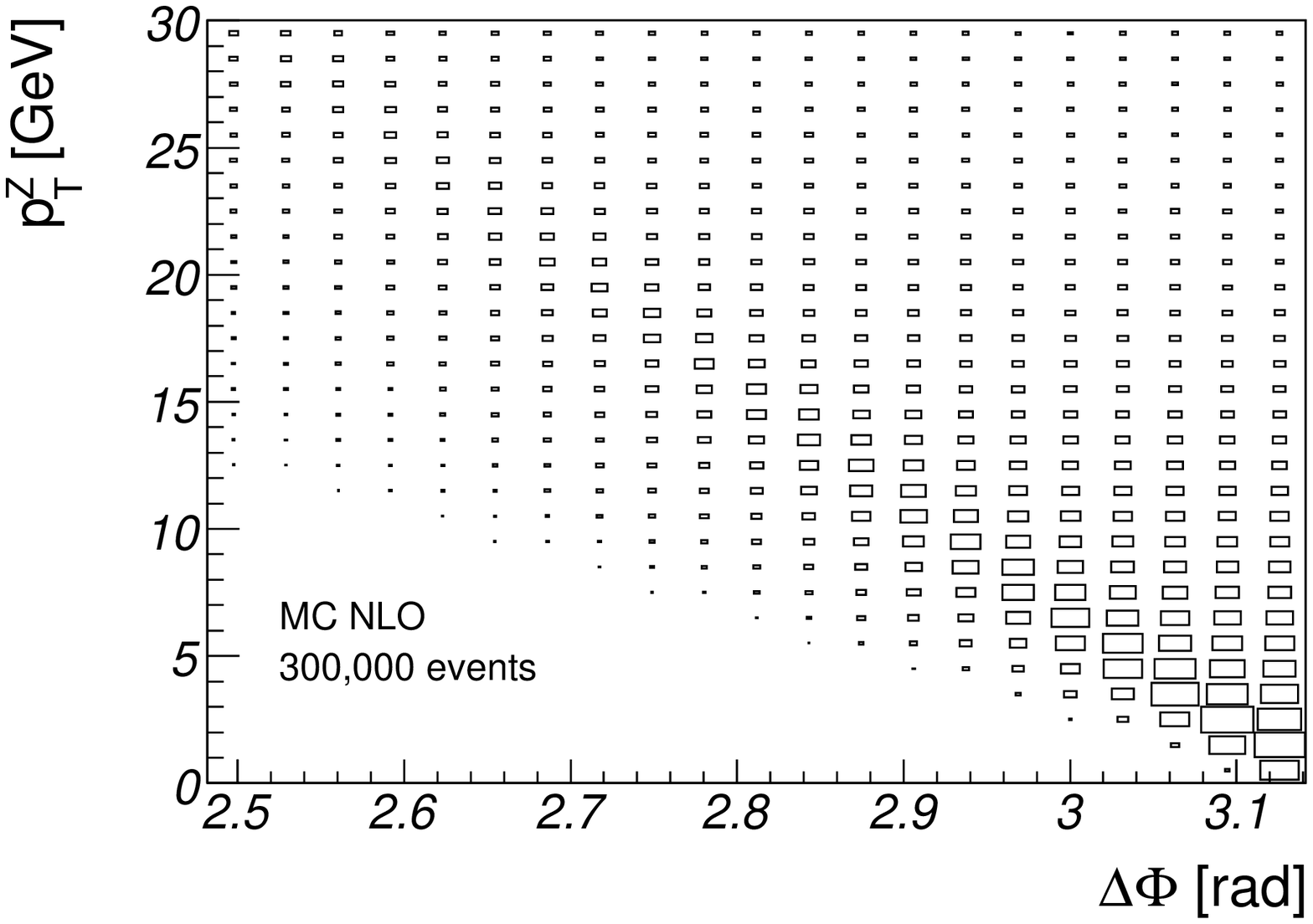}
  \caption{\label{labelAzivsPt}Distribution of the opening-angle
    between the two leptons and the $p_T$ distribution
    of the Z boson.} 
  \end{minipage}
\end{figure}

\noindent The integrated  azimuthal angular $\Delta \phi$ distribution
prediction for \textsc{MC@NLO} is shown in Figure~\ref{labelAziFull},
requiring that both decay leptons have a transverse momentum above
$20$~GeV and a pseudo-rapidity $\eta$ smaller than $2.5$, as generic
acceptance cuts applied by the LHC experiments ATLAS and CMS. The
corresponding $\Delta \phi$ distribution vs. the transverse momentum
distribution of the Z is shown in Figure~\ref{labelAzivsPt}. It can be 
interpreted as a matrix $M$, relating a given $p_T$ distribution to
a $\Delta \phi$ distribution, via 

\begin{equation}
\label{eqnMatrix}
\Delta \phi_i = M_{ij}\cdot {p_T}_j
\end{equation}
where $i$ and $j$ are the numbers of bin of the $\Delta \phi$ and $p_T$ distribution. It is required that each row of $M$ is normalized to unity. The basic idea is to use the measured $\Delta\phi$ distribution to estimate the $p_T$
distribution, exploiting their relation through the matrix $M$. The inverse matrix $M^{-1}$ directly relates  $p_T$ distribution to the measured $\vec{\Delta\phi}$ by a simple matrix multiplication. The matrix $M$ has significant off-diagonal entries and a non-uniform distribution of values on the diagonal. Hence, the inversion of $M$ and the statistical fluctuations in $\vec {\Delta \phi}$ induce large fluctuations in $\vec {p_T}$ and therefore sub-optimal results.\\

\noindent Therefore it was chosen to adjust the MC predicted $p_T$
distribution iteratively to minimize the difference in the corresponding
${\Delta\phi}_{MC}$ distribution, which is calculated via Equation~\ref{eqnMatrix}, and the measured ${\Delta\phi}_{Data}$. The difference is expressed as a $\chi^2$-value, i.e. 
\begin{eqnarray}
\label{eqnbasis}
\chi^2 &:=& \sum_i \frac{{\Delta\phi}_{Data,i} -
  {\Delta\phi}_{MC,i}}{\sigma^2_{MC,i}} \\ 
&=&  \sum_i \frac{{\Delta\phi}_{Data,i} - M_{ij}\cdot
  {p_{T,j}}}{\sigma^2_{MC,i}} 
\end{eqnarray}
It was assumed that the statistical uncertainty $\sigma$ is purely due to the measurement, as the Monte
Carlo based observables can be theoretically defined with infinite
statistics. The parameters to be varied in Equation~\ref{eqnbasis} are
all entries of the $p_T$ distribution, i.e. in realistic scenarios
more than 30. This relatively large number of free fitting parameters
dramatically hinders the minimization of Equation~\ref{eqnbasis},
especially when the statistics of the measured $\Delta\phi$
distribution is limited. Hence it was chosen to use $f(p_T,
\sigma_{PS}, \alpha, n)$, defined in Equation~\ref{eqnpara}, to model
the $p_T$ distribution, i.e. using only three free parameters
$\sigma_{PS}, \alpha$ and $n$ during the $\chi^2$ -minimization
procedure. It has been shown in Section~\ref{param} that the
parameterization of the $p_T$ spectrum  via Equation~\ref{eqnpara}
provides an adequate description up to statistics of at least $5.10^4$
events, which we assume in the present analysis.

%% file: expectedPrecision.tex
As already mentioned in Section~\ref{sec:intro}, a prominent systematic
uncertainty in the standard measurement of the
$p_T$ is induced by the uncertainty on the decay lepton momentum
measurement. In contrast, the presented approach has 
only a very weak dependence on the momentum measurement, but relies on
the measurement of $\Delta \Phi$, which has in general extremely high
precision in most collider experiments. The expected precision of
modern detectors, like ATLAS or CMS, is a magnitude smaller than the required binning in the
$\Delta\Phi$ distribution. For this reason, the $\Delta\Phi$ distribution is not required to be unfolded, which is an additional
advantage compared to the standard measurement.  \\

\noindent The uncertainties on the lepton transverse momentum
measurement can be parameterized to first order as 

\[p_T \rightarrow a\cdot p_T + \mathrm{Gauss}(m,\sigma)\]

\noindent where $a$ is a scale parameter, $m$ is an offset parameter of and
$\sigma$ is the width of an additional Gaussian (resolution)
uncertainty. The scale parameter $a$ is assumed to have an uncertainty
of $1\%$, the resolution parameter $\sigma$ is assumed to have an
uncertainty of 300~MeV. Moreover, we assume a systematic shift of
$2\times10^{-3}rad$ on the $\Delta\Phi$ measurement. The latter
assumption is rather conservative, keeping in mind the precise
$\Delta\Phi$ resolution.  \\

\noindent With this assumptions of systematic uncertainties, we can compare both measurement techniques. Table \ref{TableLabelTheoRatios2} shows the comparison of function parameters $\alpha,
\sigma, n$ of Equation~\ref{eqnpara} and its maximum $max$ for different
measurement methods and systematic uncertainties. \textit{Standard}
denotes that the parameter values have been obtained by directly
fitting the predicted transverse momentum distribution. $\Delta\Phi$
denotes that the corresponding values have been obtained with fitting
the opening angle distribution. \textit{Ideal} labels that a perfect
detector has been assumed, i.e. with perfectly known resolution, while
\textit{distorted} assumes the stated uncertainties on the detector
resolution. The given values are based on 500.000 selected Z boson
events in a specified leptonic decay channel, generated with
\textsc{MC@NLO}. \\

\noindent The parameters in the ideal, standard column are the
reference values for the comparisons. The values of the $\Delta\Phi$
approach agree within their statistical uncertainties to the reference
values. Moreover, the assumed systematic uncertainty on the
$\Delta\Phi$ measurement has no significant effect of the fit
results. This is not the case for the distorted standard measurement,
where a significant difference compared to the ideal measurement can
be observed. The statistical precision of the $\Delta\Phi$-approach is
reduced less, by a factor of $2$. Such a decrease is expected due to
the integration over rapidity and the mass of the Z boson and hence
not all statistical information used. This can be partially recovered
when taken the rapidity information into account during the fitting
procedure, or when restricting the analysis to a smaller rapidity
range, e.g. $|y|<1.0$. Nevertheless, the systematic difference between
the standard measurement is worse compared to the statistical
uncertainty of the $\Delta\Phi$-approach. 

\begin{table}[h]
\begin{small}
\begin{center}
\begin{tabular}{l|c|c|c|c}
\hline \hline
Parameter	& standard & standard measurement	& $\Delta\Phi$
approach	& $\Delta\Phi$ approach \\ 
			& (ideal) 	  & (distorted)	& (ideal)
& (distorted) \\ 
\hline
$max$ 		&	3.89 (0.02)	& 4.04 (0.02)	& 3.90 (0.04)
& 3.90 (0.04) \\ 
\hline
$\sigma$ 		&	3.25 (0.03)	& 3.10 (0.05)	& 3.22
(0.8) & 3.23 (0.09) \\ 
\hline
$\alpha$ 		&	0.90 (0.01)	& 0.85 (0.01)	& 0.89
(0.02) & 0.91 (0.03) \\ 
\hline
$n$ 			&	3.57 (0.02)	& 3.94 (0.04)	& 3.58
(0.06) & 3.59 (0.06) \\ 
\hline
\end{tabular}
\caption{Comparison of function parameters $\alpha, \sigma, n$ and the
  maximum of the function for different measurement methods and
  systematic uncertainties. The statistical uncertainties are given in
  parenthesis.} 
\label{TableLabelTheoRatios2}
\end{center}
\end{small}
\end{table}

%% file: comparisonMC.tex
\begin{figure}[htbp]
  \begin{center}
    \subfigure[]{\includegraphics[width=0.56\figwidth]{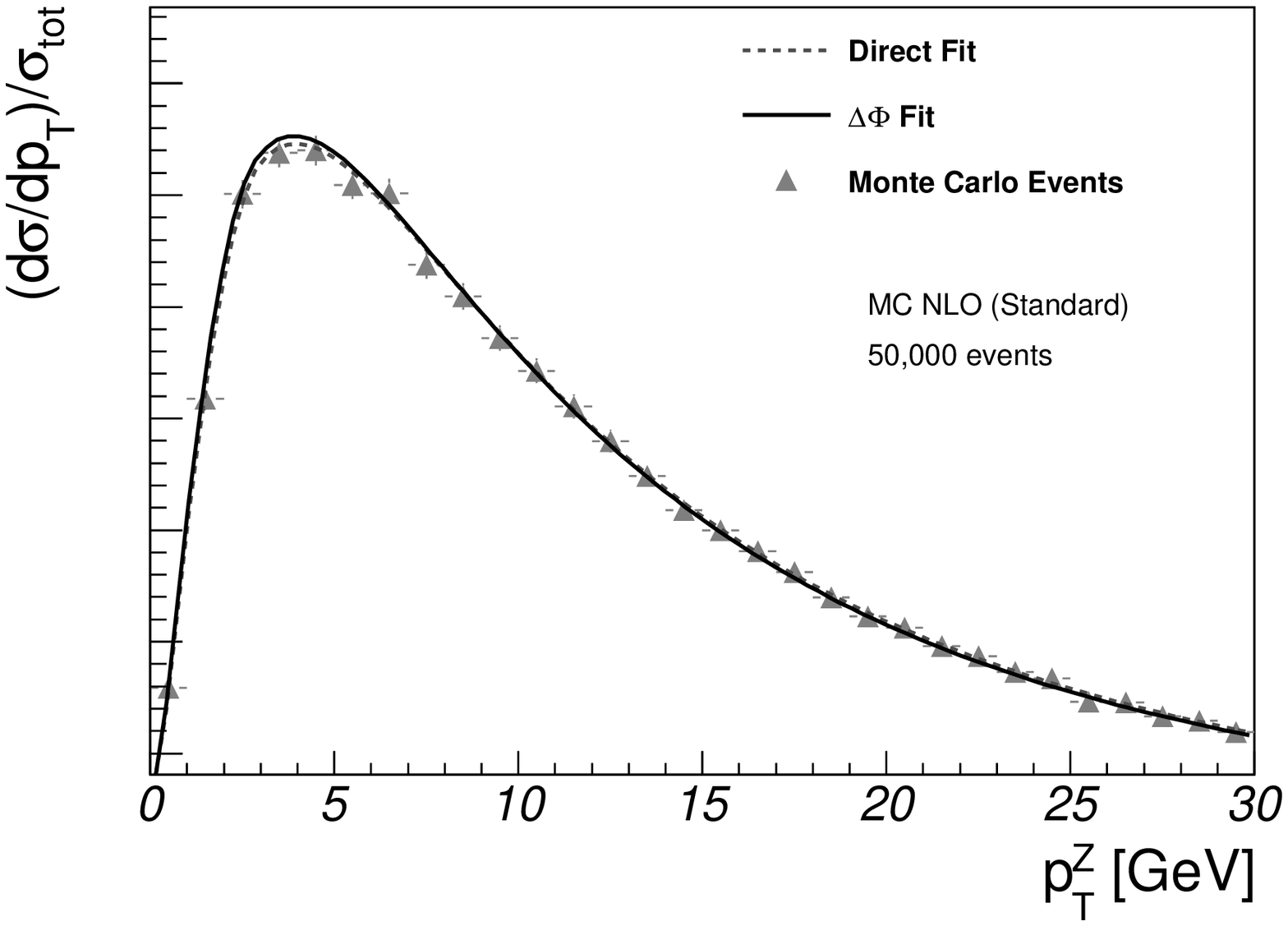}\label{Mpythia}}   
    \subfigure[]{\includegraphics[width=0.56\figwidth]{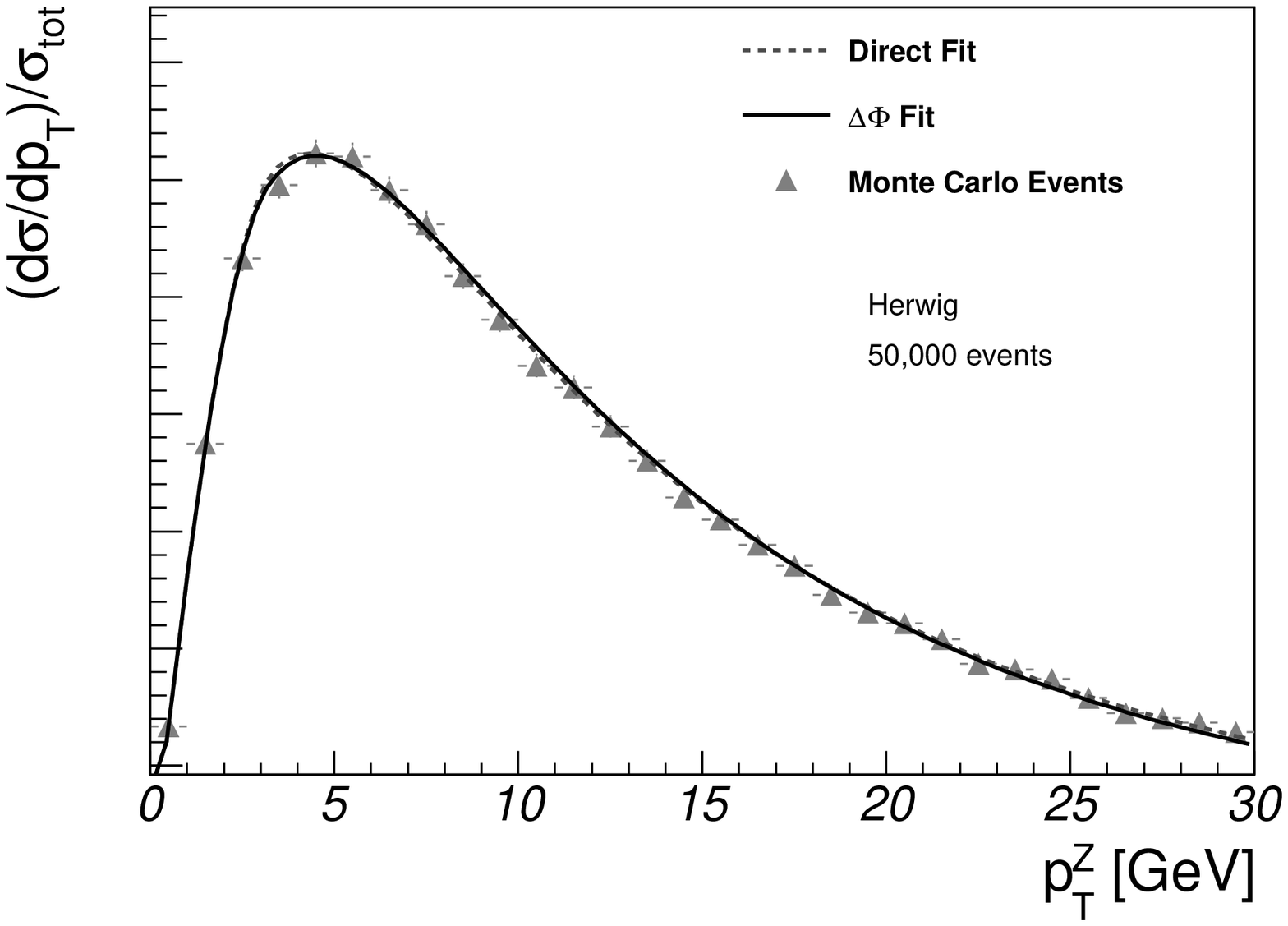}\label{Mmcatnlo}} 
    \subfigure[]{\includegraphics[width=0.56\figwidth]{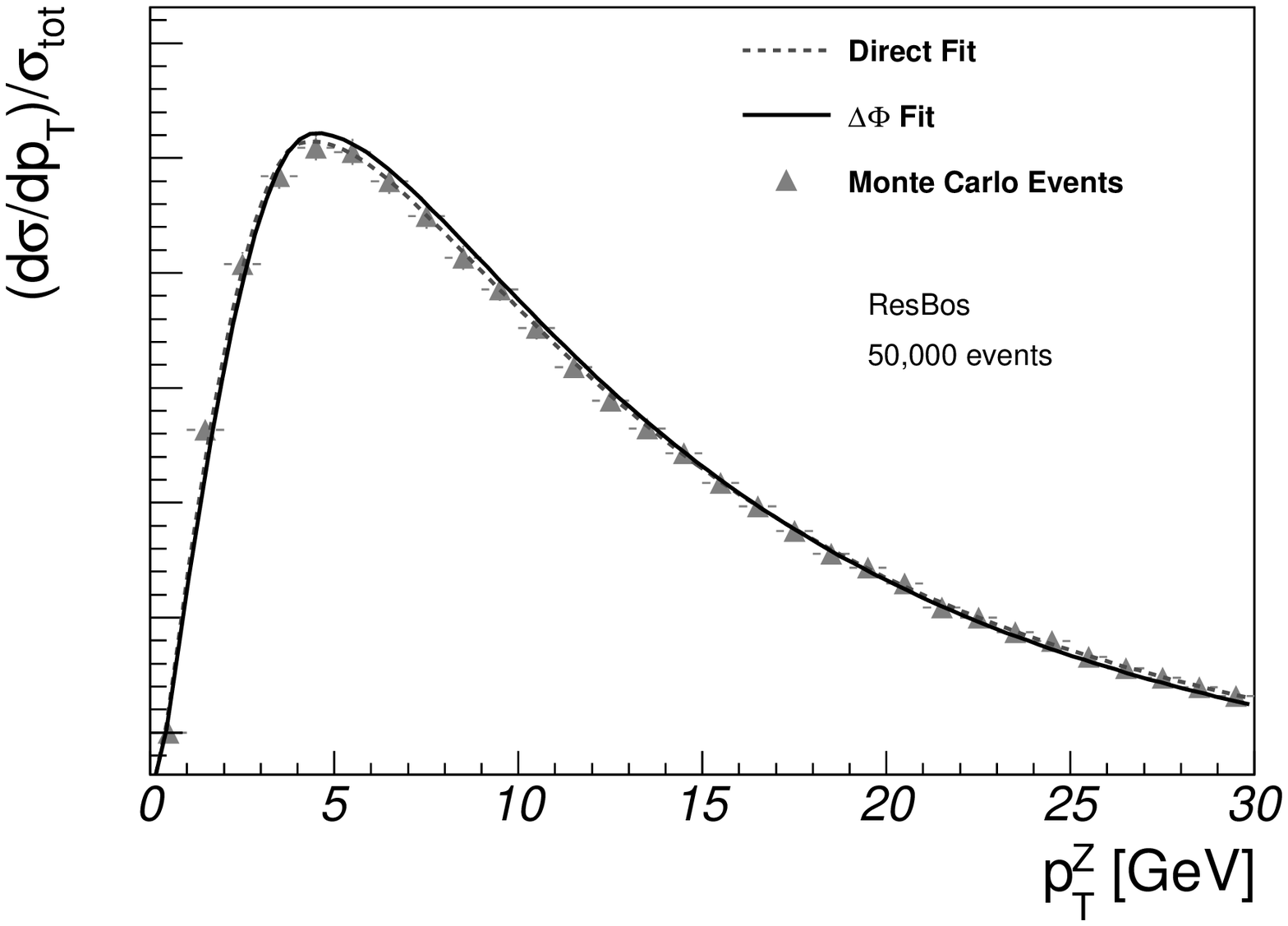}\label{MresbosStd1}}
    \subfigure[]{\includegraphics[width=0.56\figwidth]{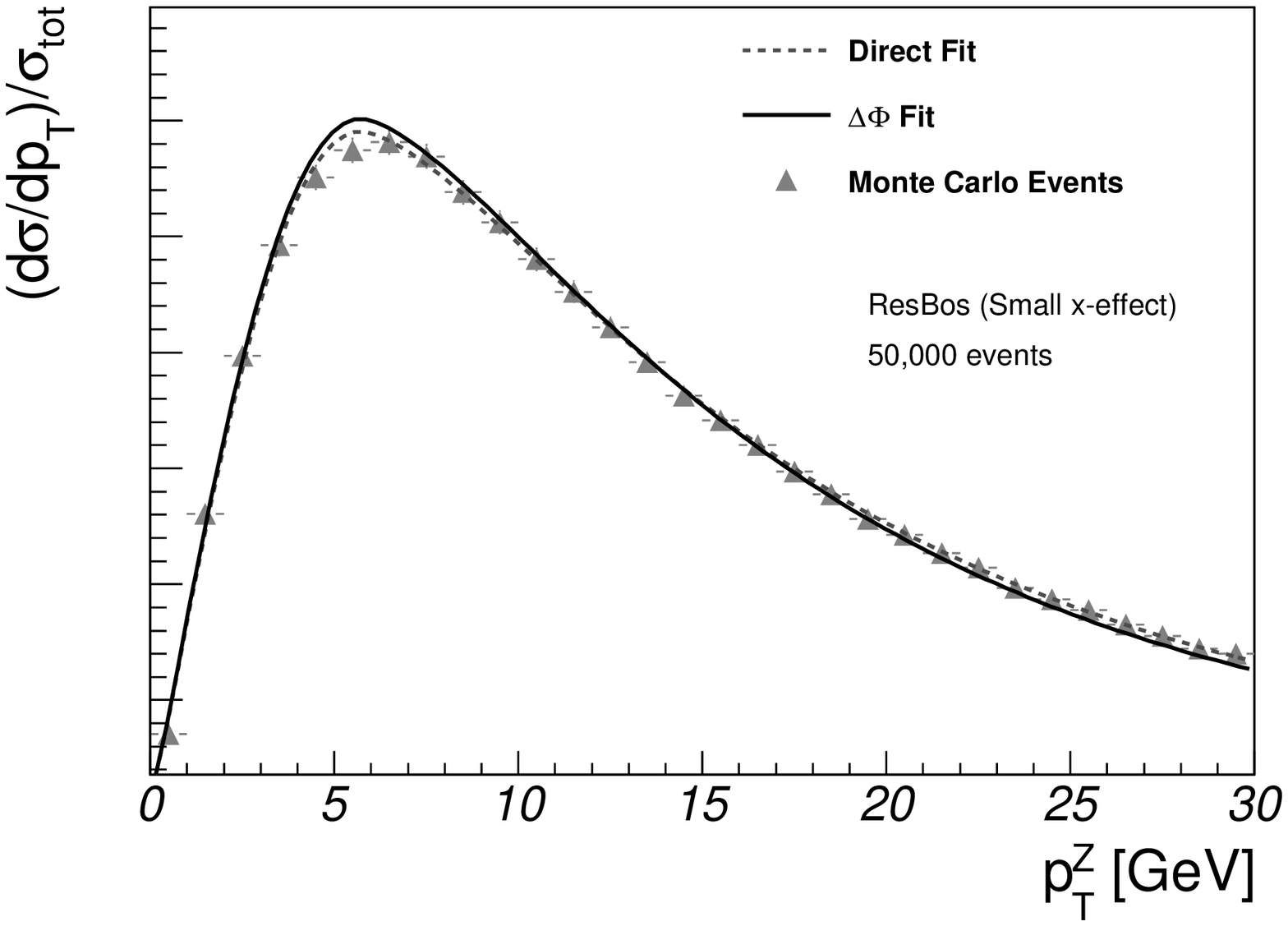}\label{MresbosBrd1}}
  \end{center}
  \caption{Fitted function for different Monte Carlo generator
    programs: {\tt MC@NLO} (a), {\tt HERWIG} (b),{\tt RESBOS}
    (c), {\tt RESBOS} with x-broading (d). The dashed function has
    been directly fitted to the Monte Carlo predicted transverse
    momentum distribution, while the solid function line has been
    obtained with by the opening angle distribution. The errors and
    therefore the quoted likelihoods correspond to 50.000
    reconstructed events.. 
  \label{labelFigureComMC}}
\end{figure}

As previously discussed, the values of the matrix $M$ are predicted by
a Monte Carlo generator. In this section, it will be discussed to
which extent the final $p_T$ measurement is independent of a
specific Monte Carlo generator program. In order to test this
independence, an attempt was made to predict the $p_T$ spectrum of one generator using the predicted matrix $M$ of a second
generator. Again, it was chosen that the matrix is based on
{\tt MC@NLO}. The $p_T$ and $\Delta \Phi$ spectra used for the
comparison and fitting, respectively, are taken from {\tt HERWIG},
{\tt RESBOS}, and {\tt RESBOS} including the x-broadening
effect. Each spectra is based on 50.000 reconstucted Z boson events,
corresponding roughly to an integrated luminosity of ${\cal L} = \int
100\,\mbox{pb}^{-1}$ at the LHC at a collision energy of $7$~TeV. The
resulting $p_T$ spectra, including the functional description obtained
with a direct fit and via the $\Delta\Phi$ measurement are shown for
the different generators in Figure~\ref{labelFigureComMC}. The
corresponding fitting parameters are shown in Table
\ref{TableLabelTheoRatios}. \\

\begin{table}[htb]
\begin{small}
\begin{center}
\begin{tabular}{l|c|c|c|c}
\hline \hline
Parameter	& {\tt MC@NLO}	& {\tt HERWIG}		& {\tt RESBOS}
& {\tt RESBOS} \\  
			& 			&
&			& (small x) \\ 
\hline
$\sigma_{Truth}$ 		&	3.31 (0.09)	& 3.80 (0.17)
& 4.27 (0.11) & 5.50 (0.13) \\ 
$\sigma_{\Delta\Phi}$ 	& 	3.14 (0.30)	& 3.70 (0.44)	& 4.45
(0.29) & 5.70 (0.28) \\ 
\hline
$\alpha_{Truth}$ 		&	0.90 (0.03)	& 0.92 (0.04)
& 0.97 (0.03) & 1.07 (0.03) \\ 
$\alpha_{\Delta\Phi}$ 	& 	0.83 (0.09)	& 0.88 (0.11)	& 0.99
(0.07) & 1.07 (0.07) \\ 
\hline
$n_{Truth}$ 			&	3.55 (0.09)	& 3.81 (0.14)
& 3.62 (0.14) & 4.04 (0.25) \\ 
$n_{\Delta\Phi}$ 		&	3.86 (0.20)	& 4.12 (0.39)
& 3.96 (0.40) & 4.84 (0.77) \\ 
\hline
$max_{Truth}$ 			&	3.95 (0.07)	& 4.34 (0.10)
& 4.48 (0.09) & 5.50 (0.13) \\ 
$max_{\Delta\Phi}$ 		&	4.11 (0.14)	& 4.50 (0.22)
& 4.70 (0.16) & 5.78 (0.20) \\ 
\hline
\end{tabular}
\caption{Comparison of function parameters $\alpha, \sigma, n$ and the
  maximum of the function for different Monte Carlo generator
  programs. The subscript \textit{truth} denotes that the function has
  been directly fitted to the predicted transverse momentum
  distribution, while the subscript $\Delta\Phi$ denotes that the
  corresponding values have been obtained with fitting the opening
  angle distribution. The values correspond to 50.000 selected
  events.} 
\label{TableLabelTheoRatios}
\end{center}
\end{small}
\end{table}

\noindent The $\Delta\Phi$ fitted values agree within their statistical
uncertainty to the values, obtained by a direct fit to the truth
$p_T$ distribution. As already mentioned in the previous section,
the statistical precision and also the systematic differences are
mainly due to the integration over the Z boson rapidity. \\

\subsection{\label{XBroadening} X-Broadening Effects}

As a final example we want to demonstrate that the presented
$\Delta\Phi$ based $p_T$ spectrum measurement can be also used to
test the x-broadening prediction at LHC for early data, i.e. small
integrated luminosities. \\

\begin{figure}[htb]
\centering
  \begin{minipage}[b]{0.56\figwidth}
  \includegraphics[width=0.56\figwidth,angle=0]{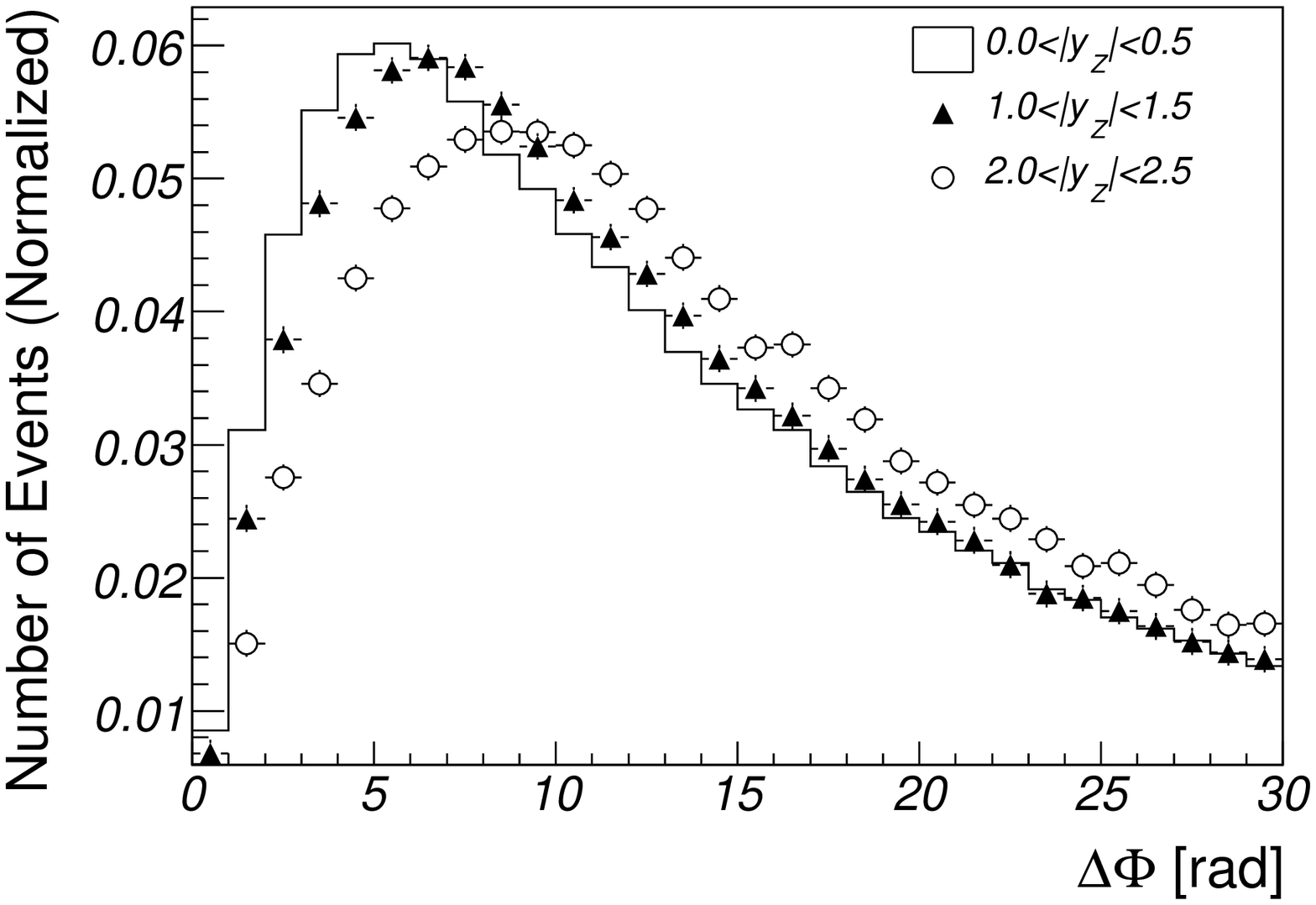} 
  \caption{\label{labelTransMass}Transverse Momentum Spectrum for
    different rapidity ranges of the Z boson}  
  \end{minipage}
  \begin{minipage}[b]{0.56\figwidth}
    \includegraphics[width=0.56\figwidth, angle=0]{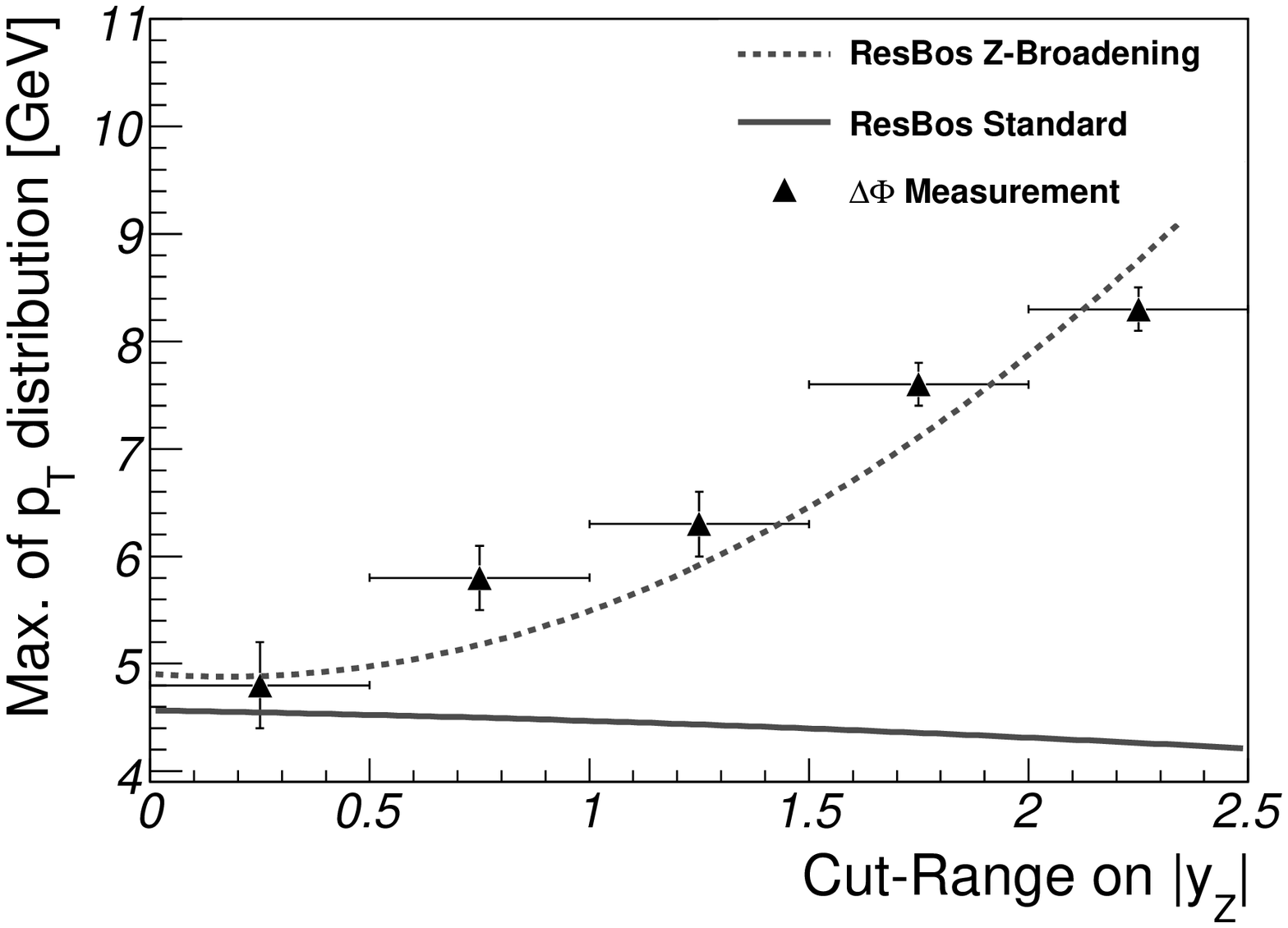}
  \caption{\label{labelXBroadening}X-Broadening determination with
    50.000 reconstructed events.} 
  \end{minipage}
\end{figure}

\noindent Instead of directly measuring the $p_T$ spectrum to test the
x-broadening effect, we propose to measure the maximum of the $p_T$
spectrum for different intervals of the Z-boson rapidity. Figure~\ref{labelTransMass} shows the predicted $p_T$ spectra for different
$y_Z$. Larger Z boson rapidities $y_Z$ test smaller x-regimes of the
interacting partons. Hence it is expected that the x-broadening
enhances for larger $y_Z$ values, i.e. the maximum the $p_T$ shifts
to larger values. The measurement of the maximum dependence of the
$p_T$ on $y_Z$ does not only allow to see a possible x-broadening
effect, but also to constrain some model parameters.  \\

\noindent To test the $\Delta\Phi$ based measurement, we again assume
a statistics of $50.000$ reconstructed Z boson decays, distributed in
five rapitity intervals ($[0,0.5]$, $[0.5, 1.0]$, $[1.0,1.5]$,
$[1.5,2.0]$ and $[2.0,2.5]$). In each interval, we perform the
$\Delta\Phi$ based fit and extract the maximum of the corresponding
$p_T$ distribution. The results are shown in Figure~\ref{labelXBroadening}. It becomes evident that we can distinguish
between the standard prediction and the small-x prediction using the
$\Delta\Phi$ based fit on a relatively small data sample. \\

%% file: perspectives.tex
\noindent The present work has two components. First, a phenomenological,
three-parameter parametrization of the heavy boson \pt\ distribution
was introduced, which proved sufficiently versatile to describe the 
available theoretical predictions. This function has many
applications, in quantifying differences between predictions and
assisting the measurement procedure.\\

\noindent Secondly, we propose a measurement method that is free of any
energy resolution systematics, while remaining a measurement of the
$p_T$ distribution. The result can thus be directly compared to
theoretical predictions of this quantity. We presented here a
simplified version of this algorithm, based on a map representing the
$\Delta\phi$ distribution at given \pt. The map was integrated over
rapidity for simplicity of this presentation, at the cost of some
statistical power. A complete treatment will have to account for the
lepton pseudo-rapidity event by event.\\

\noindent The computation of this map was performed by Monte Carlo simulations,
and relies on the well known $Z$ boson mass distribution, on
kinematics, and on the $\phi^*$ distribution in the Collins-Soper
frame, which can be computed perturbatively with good precision. The
uncertainty induced by this assumption can be expected to be small,
but will have to be quantified by further study. A realistic
measurement will need sufficient statistics, typically $O(50000)$
events, and would benefit from an analytical calculation of the map.\\

\noindent In summary, we have proposed here a method that takes as
input reliably computed quantities on the theoretical side, and
precisely measured angles on the experimental side. The resolution
of the lepton energy measurment enters only through the kinematic
selections, with negligible effect. The statistical power of the
method is about a factor two less than a direct measurement, but we
expect that the reduced systematic uncertainties involved here will
compensate for this in the long term.

%% file: acknowledgements.tex
One of the authors (M.S) gratefully acknowledges CEA/IRFU for support.

%% file: deltaphi.bbl
\begin{thebibliography}{99}

\bibitem{Collins:1984kg}
  J.~C.~Collins, D.~E.~Soper and G.~Sterman,
  Nucl.\ Phys.\  B {\bf 250} (1985) 199.


\bibitem{hep-ph/9311341}
  G.~A.~Ladinsky and C.~P.~Yuan,
  Phys.\ Rev.\  D {\bf 50}, 4239 (1994)
  [arXiv:hep-ph/9311341].


\bibitem{hep-ph/9704258}
  C.~Balazs and C.~P.~Yuan,
  Phys.\ Rev.\  D {\bf 56}, 5558 (1997)
  [arXiv:hep-ph/9704258].

\bibitem{hep-ph/0212159}
  F.~Landry, R.~Brock, P.~M.~Nadolsky and C.~P.~Yuan,
  Phys.\ Rev.\  D {\bf 67}, 073016 (2003)
  [arXiv:hep-ph/0212159].


\bibitem{hep-ph/0603175}
  T.~Sjostrand, S.~Mrenna and P.~Z.~Skands,
  JHEP {\bf 0605}, 026 (2006)
  [arXiv:hep-ph/0603175].

\bibitem{hep-ph/0210213}
  G.~Corcella {\it et al.},
  arXiv:hep-ph/0210213.

\bibitem{hep-ph/0609070}
  K.~Melnikov and F.~Petriello,
  Phys.\ Rev.\  D {\bf 74}, 114017 (2006)
  [arXiv:hep-ph/0609070].

\bibitem{arXiv:0805.2093}
  N.~Besson, M.~Boonekamp, E.~Klinkby, T.~Petersen and S.~Mehlhase  [ATLAS
                  Collaboration],
  Eur.\ Phys.\ J.\  C {\bf 57}, 627 (2008)
  [arXiv:0805.2093 [hep-ex]].

\bibitem{arXiv:0807.4956}
  M.~Vesterinen and T.~R.~Wyatt,
  Nucl.\ Instrum.\ Meth.\  A {\bf 602}, 432 (2009)
  [arXiv:0807.4956 [hep-ex]].

\bibitem{Gaiser:1982yw}
  J.~Gaiser, ``Charmonium Spectroscopy From Radiative Decays Of The J / Psi And
  Psi-Prime,'' (PhD thesis), SLAC-R-255.

\bibitem{arXiv:0812.0770}
  S.~Frixione and B.~R.~Webber,
  arXiv:0812.0770 [hep-ph].


\bibitem{hep-ph/0410375}
  S.~Berge, P.~M.~Nadolsky, F.~Olness and C.~P.~Yuan,
  Phys.\ Rev.\  D {\bf 72}, 033015 (2005)
  [arXiv:hep-ph/0410375].

\bibitem{Gribov:1972ri}
  V.~N.~Gribov and L.~N.~Lipatov,
  Sov.\ J.\ Nucl.\ Phys.\  {\bf 15}, 438 (1972)
  [Yad.\ Fiz.\  {\bf 15}, 781 (1972)].

\bibitem{Altarelli:1977zs}
  G.~Altarelli and G.~Parisi,
  Nucl.\ Phys.\  B {\bf 126}, 298 (1977).

\bibitem{Dokshitzer:1977sg}
  Y.~L.~Dokshitzer,
  Sov.\ Phys.\ JETP {\bf 46}, 641 (1977)
  [Zh.\ Eksp.\ Teor.\ Fiz.\  {\bf 73}, 1216 (1977)].

\bibitem{Collins:1977iv}
  J.~C.~Collins and D.~E.~Soper,
  Phys.\ Rev.\  D {\bf 16}, 2219 (1977).

\end{thebibliography}
